\documentstyle[twoside,aps,prl,multicol,epsf]{revtex}
\newcommand\lsim{\lower 3pt\hbox{$\buildrel < \over\sim$}}
\newcommand\gsim{\lower 3pt\hbox{$\buildrel > \over\sim$}}
\newcommand\lessgtr{\lower 3 pt\hbox{$\buildrel < \over > $}}
\begin{document}
\title{Nonlocal bottleneck effect in two-dimensional turbulence}
\author{D.\ Biskamp, E.\ Schwarz, and A.\ Celani}
\address{Max-Planck-Institut f\"ur Plasmaphysik,\\
85748 Garching, Germany }
\draft
\maketitle
\begin{abstract}
The bottleneck pileup in the energy spectrum is investigated
for several two-dimensional (2D) turbulence systems by 
numerical simulation using high-order diffusion terms to amplify
the effect, which is  weak for normal diffusion.
For 2D magnetohydrodynamic (MHD) turbulence, 2D electron MHD (EMHD) turbulence
and  2D thermal
convection, which all exhibit  
direct energy cascades, a nonlocal behavior is found
resulting in a logarithmic enhancement of the spectrum.

\end{abstract}


\begin{multicols}{2}

The local enhancement of the energy spectrum 
in front of the dissipation range, which is now generally called
bottleneck effect, is a well-established phenomenon. 
It has been
observed in numerous experiments [1],[2] and numerical simulations
[3],[4],[5], and has been discussed theoretically [6] pointing
out the physical mechanism. Even a quantitative formula
was derived
assuming a Batchelor fit for the second order structure function
[7],[8]. It is interesting to note that the magnitude of the bottleneck 
effect, however, seems to depend on the character of the turbulent eddies. 
In simulations of supersonic turbulence [9] the irrotational compressible
part of the velocity field exhibits a considerably 
weaker spectral enhancement 
than the solenoidal part. For higher-order dissipation terms
$\nu\nabla^2 \rightarrow -\nu_n(-\nabla^2)^n$, as often used in
turbulence simulations to maximize the inertial range, the amplitude
of the bottleneck effect increases, such that for $n\gg1$
it seems  to affect also the low-$k$  inertial range behavior [10],
though these results are probably not asymptotic.

Contrary to the attention the bottleneck effect attracted in 
three-dimensional (3D) turbulence, it has to our knowledge not yet been
discussed in 2D turbulent systems. It is true that 
for the enstrophy cascade in 2D Euler turbulence
no such effect exists, which is attributed to the negative
sign of the eddy viscosity [11], making the energy spectrum
slightly steeper than the corresponding Kolmogorov law, 
$E_k \sim k^{-3}(\ln k/k_0)^{-1/3}$. But for
2D systems dominated by a direct energy cascade there is no a priori
argument, why the same mechanism leading to the bottleneck effect
in 3D should not also be active in 2D. 
The effect seems, however, to be much weaker, since 
numerical simulations of such turbulent systems, in particular
in 2D magnetohydrodynamics (MHD) [12],[13],[14] and 2D electron 
magnetohydrodynamics (EMHD) [15],[16] 
have found energy spectra 
exhibiting almost perfect power laws down to the
dissipative fall-off with no visible bottleneck pileup.

One could argue that the difference is only due to the geometry of
the triad interactions,  now restricted to one plane, and
that applying the analysis of Ref.\ [8] to a 2D system might lead
to a less pronounced effect than in 3D. 
The only change in the
algebra is to replace the integral in the expression for $E_k\propto
\int^{\infty}_0kr \sin krD(r)dr$, see [8], 
by $\int^{\infty}_0krJ_0(kr)D(r)dr$, where $D(r)=
\langle [v_r({\bf x}+{\bf r}) -v_r({\bf x})]^2\rangle $ 
is the second order longitudinal
structure function. Assuming again a Batchelor fit for $D(r)$,
a straightforward evaluation
shows that the bump on the spectrum would be of similar
magnitude and width as in the 3D case,
contrary to the much weaker effect revealed in 2D simulations, 
which invalidates
this assumption.
It is therefore necessary to investigate the character of the 
transition from the inertial to the dissipation range in 2D
turbulence more in detail.

In this letter we present results of a series of simulation runs
for the turbulence systems mentioned above, 
2D MHD, 2D EMHD,  and also 2D thermal convection [18], 
using high $n$ in order to amplify the inherently weak 2D bottleneck
effect and going to higher spatial resolution than done previously.
All three
systems are two-field models which, though formally of similar 
structure, exhibit rather
different turbulence properties. 
Here the main interest is, however, not in the physics described by these
models, for which we refer to the original papers.
We first consider EMHD turbulence,
which is most closely related to (3D) Navier-Stokes turbulence.
The 2D EMHD equations are [15]
\begin{eqnarray}
(\partial_t + {\bf v}_e\cdot\nabla)(\psi-d^2_ej)
&=& -\eta_n(-\nabla^2)^n\psi,
\label{1a}\\
(\partial_t+{\bf v}_e\cdot\nabla)(\phi-d^2_e\omega) 
+{\bf B}\cdot\nabla j &=& -\eta_n(-\nabla^2)^n\omega,
\label{1b}
\end{eqnarray}
where the flux function $\psi$ describes the magnetic field in the
plane, ${\bf B} ={\bf e}_z\times\nabla\psi,\, j=\nabla^2\psi$,
and the stream function $\phi$ describes the electron flow in the plane,
${\bf v}_e = {\bf e}_z\times\nabla\phi,\, \omega = \nabla^2\phi$. 
${\bf v}_e$ is proportional to the current density in the plane,
such that $\phi$ gives the out-of-plane field fluctuation, $\phi = 
\delta B_z$. The equations are written in nondimensional form and
$d_e =c/\omega_{pe}L$ is the normalized 
collisionless electron skin depth, for details
see [15]. The equations are solved on a periodic box of linear size
$2\pi$ using a standard pseudo-spectral method with dealiasing
according to the 2/3 rule. The dissipation terms are integrated
exactly. As in [15] we consider turbulence decaying from a random
initial state.
It has been shown in [15] that for large 
wavenumbers $kd_e>1$, 2D EMHD turbulence exhibits
a Kolmogorov energy spectrum $E_k \sim \epsilon^{2/3}k^{-5/3}$.

\begin{figure}
\epsfxsize=9truecm
\epsfbox{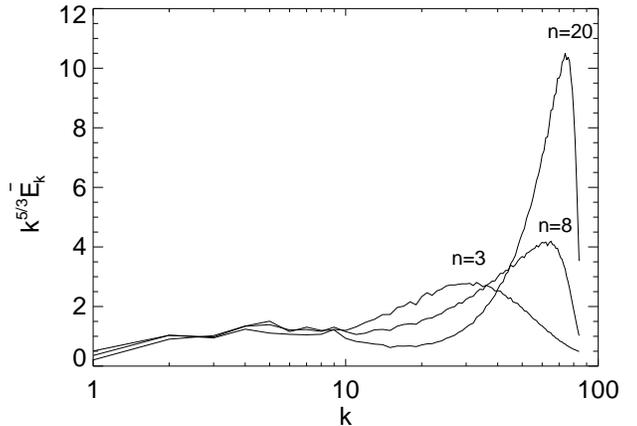}
\caption{Compensated  energy spectra $\overline{E}_k$ for 3D
EMHD turbulence simulations with $n=3,\,8,\,20$.
Normalization is such that the horizontal parts of the
spectra coincide.}   
\label{f2}
\end{figure}

\narrowtext
\begin{figure}
\epsfxsize=9truecm
\epsfbox{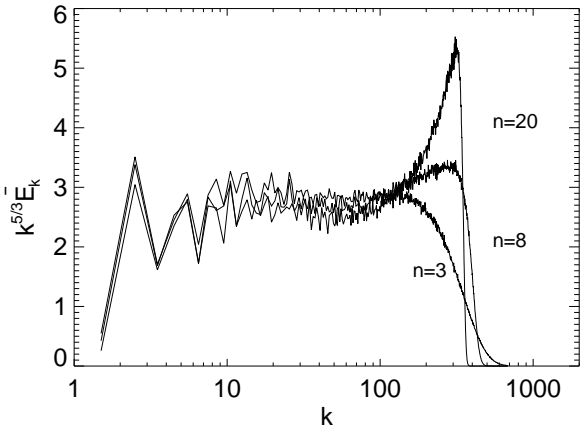}
\caption{Compensated  energy spectra $k^{5/3}\overline{E}_k$
of 2D EMHD turbulence for diffusion operator order $n=3,\,8,\,20$.
Note the linear vertical scale.}
\label{f1}
\end{figure}
Figure \ref{f1} gives the energy spectrum
$\overline{E}_k = \langle E_k
\epsilon^{-2/3}\rangle_t$
averaged over about one energy decay time.(Since at high $n$ the dissipation
length is essentially independent of $\epsilon$, the average can be
performed at constant $k$ [17].) 
Shown are 
three cases with $d_e = 0.3,\, N^2 = 2048^2,\, k_{\rm max} = 682$,
chosen as in [15],
(a) $ n=3,\, \eta_3 = 6\times10^{-11}$, (b) $n=8,\,\eta_8=10^{-38}$,
(c) $n=20,\, \eta_{20} = 10^{-96}$. The modal energy is defined
by $E_k = \sum_{\rm angle}(k^2|\psi_k|^2 + |\phi_k|^2)(1+d^2_ek^2)$. While
no bottleneck effect is visible for $n=3$, in agreement with the
spectrum shown in [15], there is a clear spectral enhancement for $n=8$ 
of $20\%$ and
for $n=20$ of about a factor of 2. 
For comparison with the corresponding 3D behavior several 3D EMHD simulation
runs have been performed for similar parameter values, though at lower
Reynolds numbers.
3D EMHD follows the equation
\[
\partial_t({\bf B} - d^2_e\nabla^2{\bf B})
- \nabla\times[{\bf v}_e\times
({\bf B} - d^2_e\nabla^2{\bf B})]
\]
\begin{equation}\hspace{3cm}
= -\eta_n(-\nabla^2)^n{\bf B},
\label{2}
\end{equation}
where ${\bf v}_e = -\nabla\times{\bf B}$.
In Fig.\ \ref{f2} we plot the compensated energy spectrum
from three simulation runs of decaying 3D turbulence
with $N^3 = 256^3,\, k_{\rm max} = 85$,  
for $d_e = 1$ and $n=3,\,8,\,20$, which
show bottleneck enhancement factors of 2.5,\,4,\,10, respectively.
Hence the bottleneck effect is indeed quantitatively
much weaker in  2D  than in 3D. 
(Note that for $kd_e\gg 1$ 3D EMHD formally reduces to the Navier-Stokes
equation in the vorticity form $\nabla^2{\bf B} = \nabla\times{\bf v}_e 
\rightarrow
\nabla \times {\bf v}$. 
In fact the $n=8$ spectrum in Fig.\ \ref{f2} is practically
identical with that observed for the corresponding Navier-Stokes case [17].)

\begin{figure}
\epsfxsize=9truecm
\epsfbox{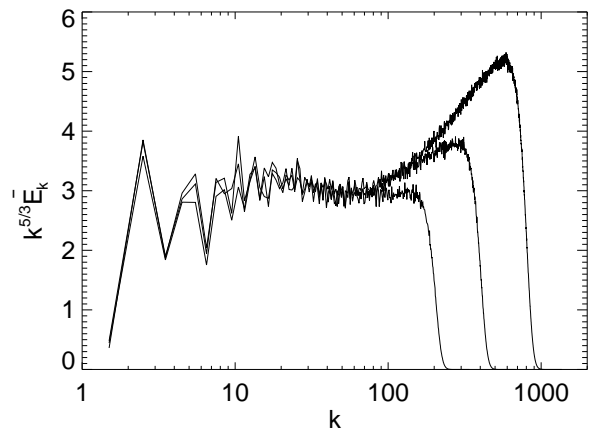}
\caption{Compensated energy spectra for three 2D EMHD turbulence
simulations with $n=8$, $\eta_8 = 10^{-32}, 10^{-36},10^{-40}$.}
\label{f3}
\end{figure}

The general understanding of the bottleneck seems to be
that the spectral enhancement, while  depending on $n$,
is independent of the extent of the inertial range. Increasing
the Reynolds number should only shift the bump to larger $k$,
but not increase its amplitude, see e.g.\ [8]. We
\noindent
 will now show that
such behavior is in general not true for 2D turbulence. Figure \ref{f3}
gives the energy spectrum for three 2D EMHD turbulence runs, where
we choose $n=8$ to amplify the effect,
$\eta_8 = 10^{-32}, 10^{-36},10^{-40}$ using
$N^2 = 1024^2,2048^2, 4096^2$, respectively. While there is
no visible bottleneck effect for $\eta_8 = 10^{-32}$, it
becomes more and more pronounced with decreasing 
dissipation coefficient leading to a flattening of the spectrum
$E_k$ in an increasingly larger fraction 
of the inertial range, $k>\kappa \sim 60$.

\narrowtext

\begin{figure}
\epsfxsize=9truecm
\epsfbox{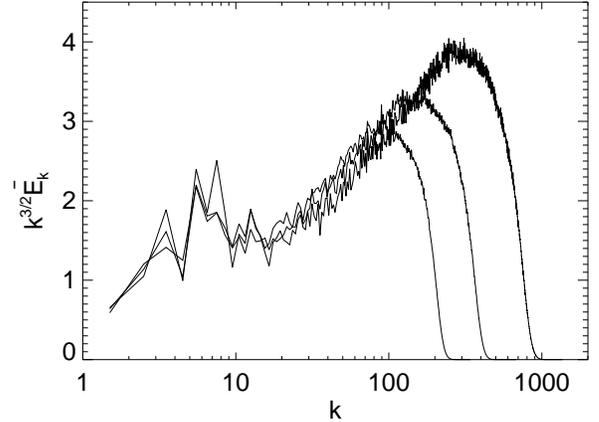}
\caption{Compensated 2D MHD energy spectra 
$k^{3/2}\overline{E}_k$ for 
$\eta_8 = \nu_8 =10^{-36}, 10^{-40}, 10^{-45}$.}
\label{f4}
\end{figure}

This behavior is not limited to 2D EMHD turbulence, but is 
found to occur in a
similar and even clearer form in 2D MHD turbulence simulations. 
Here  the dynamical equations are
\begin{eqnarray}
\partial_t\psi + {\bf v}\cdot\nabla\psi &=& -\eta_n
(-\nabla^2)^n\psi,
\label{3a}\\
\partial_t\omega + {\bf v}\cdot\nabla\omega - {\bf B}\cdot\nabla j
&=& -\nu_n(-\nabla^2)^n\omega,
\label{3b}
\end{eqnarray}
where ${\bf v} = {\bf e}_z\times\nabla\phi$ is the plasma flow
(note that in spite of the formal similarity EMHD does {\it not}
converge to MHD for $d_e \ll 1$. EMHD is limited
to $d_e>\sqrt{m_e/m_i}$, while MHD is only valid at larger, macroscopic
scales, see [15]). Previous numerical studies of decaying 2D MHD
turbulence have revealed the spectral law $E_k \sim 
(v_A\epsilon)^{1/2}k^{-3/2}$ (see e.g.\ [14]), 
where $v_A = B/\sqrt{4\pi \rho}$ is
the Alfv\'en speed and the modal energy is  $E_k =
\sum_{\rm angle}k^2(|\psi_k|^2+|\phi_k|^2)$. 
The $k^{-3/2}$ spectrum results from the
Alfv\'en effect [19],
the coupling of small-scale velocity and magnetic field fluctuations
by the magnetic field of the large-scale eddies. 
For normal diffusion no bottleneck is discernable in the energy
spectra [12], [14].
To investigate this point more closely, we choose again a high-order
dissipation operator in order to amplify the bottleneck effect,
which may be hidden in the noise level for  $n=1$.
Three simulation runs have been performed for decaying MHD turbulence
using the same numerical scheme as in the EMHD simulations described above
and similar initial conditions (called B-type in [12]).
Figure \ref{f4}
gives the time-averaged MHD energy spectra 
$\overline{E}_k = \langle E_k(v_A\epsilon)^{-1/2}\rangle_t$
plotted in compensated form
for $\eta_8 = \nu_8 =10^{-36}, 10^{-40}, 10^{-45}$ with resolutions
$N^2 = 1024^2, 2048^2, 4096^2$, respectively. Comparing with the
corresponding 2D EMHD cases given in Fig.\ \ref{f3}, Fig.\ \ref{f4} shows  
a similar qualitative trend, a nonlocal influence of the dissipation
range on the inertial range. The difference is probably due to the choice
$d_e=0.3$ in the EMHD runs in Fig.\ \ref{f3}. Since a pure scaling behavior
exists only for $kd_e>1$, the effective scaling range is shorter
by a factor 3-4 for the same resolution, such that amplitude of the
bottleneck pileup in the highest resolution EMHD case $4096^2$
in Fig.\ \ref{f3} corresponds to the lowest-resolution MHD case $1024^2$
in Fig.\ \ref{f4}.

There is a nearly linear increase of the compensated spectrum in the log-linear
plot, which suggests that for sufficiently large $k>\kappa \sim 20$
the inertial range is modified by a logarithmic factor
\begin{equation}
E_k \sim k^{-3/2}\ln (k/\kappa),
\label{4}
\end{equation}
where the magnitude of the effect is expected to depend on $n$,  
such that for $n=1$ it becomes invisibly small
at the achievable spatial resolution. The possibility of a
logarithmic factor has been discussed in [6] for 3D turbulence,
though only as a subdominant effect in the spectral correction term.
The wavenumber $\kappa$ is connected with some structure of the macro-state
of the system. The fact that $\kappa$ is lower in the MHD runs than
observed in the EMHD runs is due to the choice $d_e=0.3$ in the
latter.

These results show that the bottleneck effect in the 2D turbulence
systems considered, though very weak for normal diffusion $n=1$, exhibits a
nonlocal behavior when
\narrowtext
\noindent
enhanced by choosing high $n$.
The mechanism for this property 
must be connected with a stronger direct interaction of small- and large-scale
modes in 2D than in 3D, which is also reflected in configuration space
by the large-scale intermittency typical for 2D turbulence. 
The observed behavior is probably not due to the Alfv\'en effect,
since the latter is not present in EMHD [15].

We would like
to discuss briefly also a third type of turbulence,
2D thermal convection in the Boussinesq approximation described by  the equations
\begin{eqnarray}
\partial_tT+{\bf v}\cdot\nabla T +\partial_y\phi
&=& -\chi_n(-\nabla^2)^nT,
\label{5a}\\
\partial_t\omega+{\bf v}\cdot\nabla\omega +\partial_yT &=&
-\nu_n(-\nabla^2)^n\omega,
\label{5b}
\end{eqnarray}
again written in nondimensional form, see [18],
whre $T$ is the temperature fluctuation and ${\bf v} ={\bf e}_z\times
\nabla\phi$ the fluid velocity. 
Contrary to MHD or EMHD, where stationary turbulence 
can only be achieved by
an external stirring force, this system is linearly unstable
over a broad $k$-range with growth rate $\gamma \propto k_y/k$,
which generates a stationary level of turbulence.
This is caused by the frozen-in mean temperature gradient $T'_0$, 
$v_xT'_0=\partial_y\phi$ in the nondimensional form (\ref{5a}).
Hence there is no ideal energy invariant, instead one has
\begin{equation}
\frac{d}{dt} \int{\textstyle \frac{1}{2}}(T^2+v^2)d^2x = 2\int v_xT\,d^2x 
- \epsilon,
\label{6}
\end{equation}
where $\epsilon$ is the energy dissipation
rate. 
Equations (\ref{5a}),(\ref{5b}) have recently been studied
numerically on a periodic box using a similar scheme as in the EMHD
simulations. In spite of the anisotropic linear drive the spectra
$E^T_k = |T_k|^2$ and $E^V_k = |v_k|^2$ are highly isotropic, which
demonstrates the strong influence of the nonlinear terms.
The energy dissipation rate $\epsilon =\epsilon^V_L + 
\epsilon^T_L + \epsilon^V_s + \epsilon^T_s$ 
in (\ref{6}) consists of the dissipation on the velocity and the 
temperature fluctuations both at large ($L$) and small ($s$) wavenumbers.
The kinetic energy $v^2$ has an inverse cascade and is primarily dissipated
at small wavenumbers, where the modes are artificially damped to
prevent condensation and suppression of turbulence [18], 
$\epsilon^V_L<\epsilon^V_s$,
while the thermal fluctuation energy $T^2$ has a direct cascade, 
$\epsilon^T_L>\epsilon^T_s$.

\begin{figure}
\epsfxsize=9truecm
\epsfbox{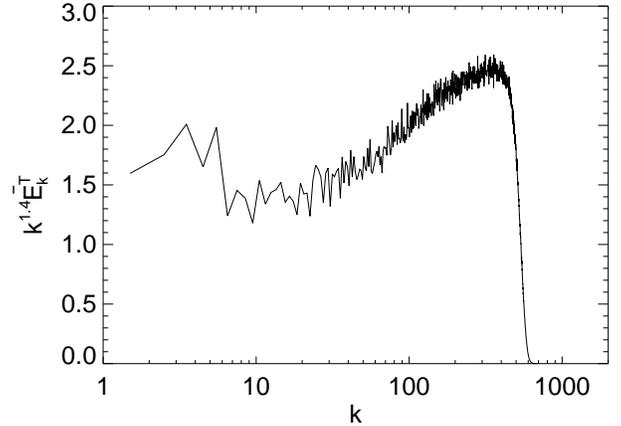}
\caption{2D thermal convection. Compensated temperature
fluctuation spectrum $k^{1.4}E^T_k$
for $n=8$.}
\label{f5}
\end{figure}

The results of [18] seem to indicate spectral laws
$E^T_k \sim k^{-1.2}$ and $E^V_k \sim k^{-2.3}$, which differ somewhat
from the expected Bolgiano scaling [20] $k^{-1.4}$ and $k^{-2.2}$,
respectively. No convincing physical argument for these deviations
could be given in [18] except the fact that the temperature
fluctuations are found to be highly intermittent, which limits the
relevance of the (nonintermittent) Bolgiano scalings $\delta_lT \sim
l^{0.2},\, \delta_lv \sim l^{0.6}$. Here we
suggest an alternative interpretation of the simulation results in [18],
namely that the deviations are caused by nonlocal effects,
which should make the spectrum of the kinetic energy with an
inverse cascade slightly steeper (cf.\ the 2D Euler case), 
while the spectrum of the temperature
fluctuations, which have a direct cascade, should be flatter than the
Bolgiano spectrum. For direct comparison with the EMHD and MHD simulations
presented above, we have performed a similar run for thermal
convection with $n=8$, $\chi_8=\nu_8=10^{-42},\, N^2=2048^2$, from which
Fig.\ \ref{f5} shows the temperature fluctuation spectrum compensated
with the Bolgiano law. The behavior is indeed very similar to
the MHD spectrum in Fig.\ \ref{f4}.

In conclusion we have investigated the spectral bottleneck pileup in
2D turbulence. While it is known that there is no such effect
in the enstrophy cascade for 2D Euler turbulence, we have  shown
the existence of the effect in 2D systems exhibiting a direct energy
cascade, in particular in MHD, EMHD and thermal convection.
The amplitude of the pileup is found to be significantly smaller than in
corresponding 3D cases even for high-order hyperdiffusion, which enforces
the tendency of spectral pileup by
making the transition between inertial range and dissipative range more
abrupt.
But contrary to the local behavior in 3D the bottleneck effect in
2D turbulence has a nonlocal character, such that the major part of the
inertial range is affected for sufficiently small dissipation coefficients.
The simulation results suggest a logarithmic modification of the spectrum.
This behavior reflects a strong direct interaction of disparate
scales, which also manifests itself in the large-scale intermittency
typically seen in configuration space plots of 2D turbulence.
Also the deviations in the spectral laws
from Bolgiano scaling in 2D thermal convection reported previously
can be attributed to this nonlocal bottleneck effect. 
As a consequence the popular use of high-order diffusion operators in
2D turbulence simulations becomes rather doubtful. 

The authors would like to thank 
Andreas Zeiler for providing the basic version of the 3D EMHD code. A.C.
\narrowtext
\noindent
has been supported
by the European Community TMR grant ERBFMICT-972459 and by
INFM (PRA TURBO).

\end{multicols}
\end{document}